\documentclass[reprint,pra, twocolumn]{revtex4}
\usepackage{textcomp}
\usepackage{amsmath}
\usepackage{amsfonts}
\usepackage{natbib}
\usepackage{tensor}
\usepackage{calc}
\usepackage{braket}
\usepackage{subfig}
\usepackage{appendix}
\usepackage{rotating}

\usepackage[unicode]{hyperref}

\begin{document}   
\title{Single particle model of a strongly driven, dense, nanoscale quantum ensemble}
\date{\today}
\author{C.S. DiLoreto and C. Rangan}
\affiliation{ \\Department of Physics, University of Windsor, Windsor ON N9B3P4. Canada}   
\begin{abstract}
We study the effects of interatomic interactions on the quantum dynamics of a dense, nanoscale, atomic ensemble driven by a strong electromagnetic field.  We use a self-consistent, mean-field technique based on the pseudo-spectral time-domain method, and a full, three-directional basis to solve the coupled Maxwell-Liouville equations.  We find that interatomic interactions generate a decoherence in the state of an ensemble on a much faster timescale than the excited state lifetime of individual atoms.  We present a novel single-particle model of the driven, dense ensemble by incorporating interactions into a dephasing rate.  This single-particle model reproduces the essential physics of the full simulation, and is an efficient way of rapidly estimating the collective dynamics of a dense ensemble.



\end{abstract}

\maketitle 
\section{Introduction}

The dynamics of a dense ensemble of quantum emitters driven by an electromagnetic field is a topic of current interest and much excitement.  Experimental and theoretical research on dense collections of atoms have studied numerous effects such as superradiance~\cite{PhysRevA.75.033802, PhysRevA.93.033407, PhysRevA.95.033839}, dipole blockade~\cite{PhysRevLett.85.2208}, collective Lamb shift~\cite{Friedberg1973101}, etc.  In nano-optics, the interest is in developing the properties of hybrid systems such as quantum dots or organic dye molecules in proximity to metal nanoparticles~\cite{PhysRevLett.89.203002, PhysRevA.95.043834,Singh,Hughes,Malinovskaya}.
In all these studies, the type and strength of interactions between the quantum emitters (henceforth referred to as ``atoms") are specific to the type of phenomenon studied.

The effects of interatomic interactions in dense ensembles that are excited by low-intensity electromagnetic fields are typically studied computationally.  Large-scale simulations have shown that interatomic interactions can shift resonance absorptions in cold dense gases \cite{PhysRevLett.116.183601, PhysRevLett.112.113603}, modify spontaneous emission rates and decoherence rates \cite{PhysRevA.90.012511,PhysRevA.94.033848,JETP112.246,PhysRevA.95.033839}, and affect overall scattering processes \cite{Sokolov2011}.  These large-scale simulations are computationally intensive as they require the evaluation of the interaction between numerous atoms (or lattice sites).  The best scaling that we have found in the literature is one that scales as the fourth power of the number of lattice sites in Ref.~\cite{PhysRevA.95.033839}.  A popular approximation is mean-field approximation such as the one used in Refs.~\cite{PhysRevA.75.033802,DIET,PhysRevB.95.115406,PhysRevA.89.022501}.  This indeed reduces the computational effort, however is still quite computationally intensive.  Thus, many calculations use further approximations such as short pulse-methods \cite{PhysRevB.95.115406, PhysRevA.89.022501}, or quantum basis sets that are of reduced dimension~\cite{DIET,PhysRevA.89.022501}. The first broadband or short-pulse approximation is used in scattering calculations of driven ensembles of classical dipoles \cite{PhysRevA.84.043802, PhysRevA.89.022501,lumerical,kall}.  In this methodology, a broadband short pulse illuminates the system and the scattered field is tracked and Fourier-transformed to yield an appropriate intensity spectrum.  In the latter approximation,  the quantization axis of the quantum emitter is along one direction - the polarization of the incident electromagnetic field~\cite{DIET,PhysRevA.89.022501}, or a time-independent quantization axis is used \cite{PhysRevA.95.033839}.  However, these approximations may not be able to accurately capture spontaneous emission from the ensemble, or inelastic scattering \cite{PhysRevA.84.043802}. In this study, we show that these approximations are inadequate for studying a strongly driven, dense quantum ensemble.

We examine the ensemble behaviour of a dense collection of approximately 4000 atoms, modelled as two-level quantum systems (2LS), that is driven by a strong plane wave electromagnetic field.  Each two-level atom interacts with the environment, and the interaction is modelled by a radiative decay rate ($\gamma$).  The states of the individual atoms therefore significantly affect local electromagnetic field intensities, and the local fields mediate the inter-atomic interactions.  In our methodology, we model the quantum evolution of the state of each atom, the spontaneous emission from which, in turn changes the electromagnetic field that is perceived by a neighbouring atom.  Thus, both the field propagation and the density matrix evolutions must be calculated simultaneously.  We solve both Maxwell's equations and the Liouville-Von Neumann equation concurrently using a pseudo-spectral time domain (PSTD) method discussed in Sec.~\ref{PSTD}.  This is a method based on a self-consistent, mean-field approach that is free of problematic self-interactions~\cite{PhysRevA.84.043802, PhysRevA.89.022501}.  

By examining the dynamics of the strongly-driven, dense ensemble of 2LS, we find that interatomic interactions create strong disorder in the ensemble states over a characteristic time that is much shorter than the excited state lifetime of a single 2LS.  This disorder imposes an effective lifetime for quantum scattering effects in an ensemble.  This implies that in order to understand the long-term dynamics of a driven, dense quantum ensemble, short-pulse/broadband techniques are inadequate.  The interatomic interactions also lead to excitation of atoms in directions other than the incident field polarization.  This indicates that for modelling a general ensemble of dense emitters, a full, three-dimensional state basis is required.  Our calculation therefore goes beyond the standard approximations by using a plane wave excitation, and uses a full three-dimensional state basis as discussed in Sec.~\ref{3dbasis}.  For an example case of an ensemble of 1 eV emitters, our calculation shows that there is a transient upshifting of incident photons that disappears in the steady state.  This disappearance is correlated with the onset of disorder in the ensemble-averaged quantum state of the ensemble.

We propose that the overall behaviour of these dense ensembles can be modelled by a single-particle, rotating wave approximation, solution to the Lindblad-von Neumann equation.  In this model, interatomic excitations are modelled by  introducing decoherence terms inspired by models of the Forster resonance energy transfers (FRET) process in biophysical systems. This approach allows the response of a dense quantum ensemble to be rapidly approximated with a single-atom model.  This representation also isolates the processes that are most significant in determining the optical response of a nanoscale, dense quantum ensemble to strong electromagnetic excitation. 

In Section \ref{sec:method}, we discuss the computational and numerical approach that was implemented to calculate the response of nanoscale dense ensembles driven at high intensities.  In Section \ref{sec:disorder}, we discuss, via a practical example, the effects that strong driving fields and its associated decoherence have on the ensemble-averaged quantum state of a nanospherical ensemble.  Section \ref{sec:single} describes the approximation technique in which we use a single particle model to simulate the average behaviour of the ensemble.  Lastly, Section \ref{sec:conclusion} summarizes our main conclusions and future outlook of this work.

\section{Theory and Implementation} \label{sec:method}

We model a dense ensemble of two-level atoms driven by a strong, linearly-polarized, electromagnetic field.  Though the driving field is polarized in one direction, spontaneous emission from each atom excites transitions in nearby atoms in other directions.  Each of the atoms contributes to a ``mean field'' that mediates the interactions between various quantum emitters.  This mean field in the ensemble is a spatially varying, 3D-vector.  Therefore, the dynamics of an individual quantum system involves a ground state and three excited states, one for each Cartesian direction of the atomic dipole interacting with the mean field as suggested in Ref.~\cite{PhysRevA.78.013806}. The calculation involves numerically evaluating the coupled Maxwell-Liouville equations in a computational space that includes the ensemble.  

The electromagnetic field evolves in time according to Maxwell's equations:
\begin{equation}		
\nabla \times \vec{E}(\vec{r},t)= -\mu_0 \frac{\partial \vec{H}(\vec{r},t)}{\partial t}\label{MaxwellH_bare},
\end{equation}
\noindent and
\begin{equation}		
\nabla \times \vec{H}(\vec{r},t)= \epsilon_0 \frac{\partial \vec{E}(\vec{r},t)}{\partial t} + \vec{J}(\vec{r},t), \label{MaxwellE_bare}
\end{equation}
\noindent where, $\vec{H}(\vec{r},t)$ and $\vec{E}(\vec{r},t)$ are the magnetic and electric fields respectively, and $\vec{J}(\vec{r},t)$ is the free current density.  

The quantum state of each emitter evolves in time according to the Lindblad-Von Neumann equation
\begin{equation}		
\dot{\rho}(\vec{r},t) = - \frac{i}{\hbar} [H(\vec{r},t),\rho(\vec{r},t)] - L(\rho(\vec{r},t)). \label{eq:Lindblad}
\end{equation}
\noindent In this evolution equation, the Lindblad superoperator, $L(\rho(\vec{r},t))$, models the decoherence in the system. This term is linear in the state density operator and is of the form:
\begin{equation}		
L(\rho) =\sum_{d}{\frac{\gamma_d}{2} (\sigma_d^\dagger \sigma_d \rho + \rho\sigma_d^\dagger \sigma_d - 2 \sigma_d \rho \sigma_d^\dagger) }. \label{eqn:linbladsuper}
\end{equation}
\noindent In this equation, $\sigma_d$ are the Lindblad operators, which are assumed to model spontaneous emissions from an excited state to the ground state, and $\gamma_d$ is the rate of spontaneous emission.  For the emission from $\ket{i} \rightarrow \ket{j}$, these operators would take the form $\sigma_d =\sigma_{ij} =\ket{j}\bra{i}$.  All non-allowed emissions have $\gamma_d=0$, and each allowed emission has a spontaneous emission rate determined by Fermi's Golden Rule \cite{nanopticsbook}.

The quantum states of the atoms contribute to the electromagnetic field via the free current density ($\vec{J}$), whose directional components ($\eta = x, y, z$)  can be found by \cite{PhysRevA.84.043802}:
\begin{equation}		
J_\eta(\vec{r}) = N_A \braket{\frac{\partial}{\partial t} \hat{\mu_\eta}(\vec{r}) } = N_A Tr (\dot{\rho}(\vec{r}) \hat{\mu_\eta} ),
\end{equation}
\noindent where $N_A$ is the number density of emitters, $\rho(\vec{r})$ is the density matrix of an emitter located at position $\vec{r}$, and $\hat{\mu_\eta}$ is the transition dipole moment operator corresponding to the $\eta^{\text{th}}$ Cartesian component of the dipole moment.  The transition dipole moment operator is directly related to the Hamiltonian of the atom as:
\begin{equation}		
\hat{\mu_\eta} = -\frac{\partial \hat{H}}{ \partial E_\eta}.
\end{equation}

\subsection{Generalized Directional State Basis}
\label{3dbasis}
\begin{figure*}
	\centering
	\includegraphics[width = 5in]{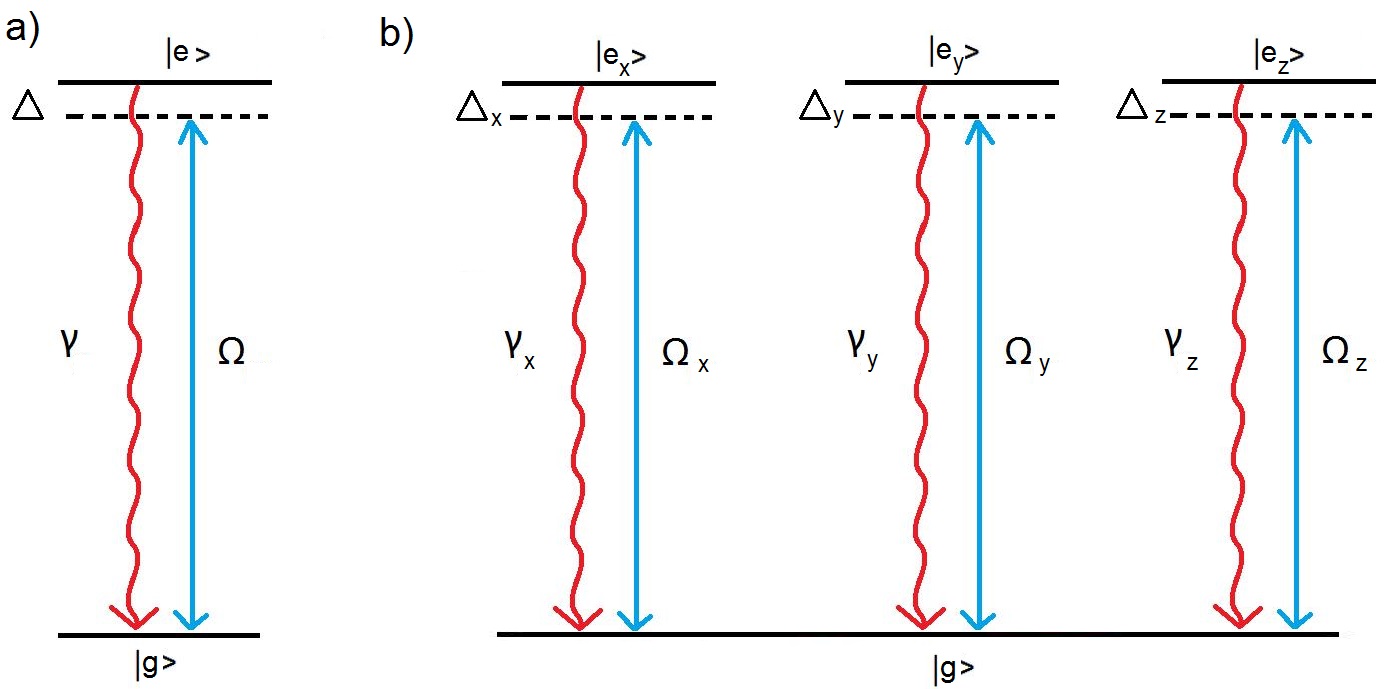}
	\caption{a) When the polarization of an electromagnetic field sets the quantization axis of an atom, the effective quantum system is a two-level system with the direction of the transition dipole oriented along that polarization direction.  b) When the polarization of an incident control field is different from the quantization axis of an atom, the effective quantum system is a four-level system with a dipole transition oriented along each field component. $\Omega$'s represent the field-atom interaction frequency, $\Delta$'s are the detuning between the frequency of the driving field and the transition frequency of the 2LS, and $\gamma$'s are the spontaneous emission rates from the excited states to the ground state.	\label{fig:DirectionalStates}}
\end{figure*}

In the quantum control of a single two-level atom by an incident electromagnetic field, the quantization axis is assumed to be along the direction of polarization, and the two atomic levels $\ket{g}$ and $\ket{e}$ are coupled with a transition strength proportional to $\mu E(\vec{r})$.  In a driven ensemble of atoms, though the driving field is polarized in one direction, spontaneous emission from each quantum system excites transitions in nearby quantum systems in other directions.  This requires the consideration of all three components of the dipole moment operator.  Rather than work in the angular momentum basis, a simpler way to approach this problem is to introduce a ``directional" state basis \cite{PhysRevA.78.013806}.  These ``directional" states are those accessed by transitions that are driven by a single field polarization  as depicted in Figure~\ref{fig:DirectionalStates}.  This results in an effective four-level system which can display quantum interference.  

The Hamiltonian of a two-level atom interactive with an electromagnetic field in this directional state basis is:
	\begin{equation}		
	H =  \left ( \begin{matrix}  
	0 & \hbar\Omega_{e_x,g} & \hbar\Omega_{e_y,g} & \hbar\Omega_{e_z,g} \\
	\hbar\Omega^*_{e_x,g}&E&0&0\\
	\hbar\Omega^*_{e_y,g} &0&E&0\\
	\hbar\Omega^*_{e_z,g} &0&0&E\\
	\end{matrix} \right ),
	\end{equation}
\noindent where the energy of the ground state is set to zero; the degenerate, excited, directional states have energy $E$, and the dipole-field interaction takes the form $\Omega_{e_\eta,g} = \frac{\mu_{e_\eta,g} E_\eta}{\hbar}$, where $\eta=(x,y,z)$.

\subsection{Mean-field Interatomic Interaction}

A microscopic representation of a large number of open quantum systems interacting with one another is computationally intensive.  Since the Lindblad-Von Neumann equation involves matrix multiplication, this computation becomes onerous for a large number of atoms in the ensemble; since even the most modern, optimized methods scale more slowly than $M^2$ \cite{davie2013improved}, where $M$ is the total number of states (for $N$ atoms, $M=4N$ for the atomic structure in Fig. \ref{fig:DirectionalStates}b).

Therefore we describe the interaction between the members of the ensemble using a mean-field method.  In this method, spatially separated atoms do not directly interact with one another through the Hamiltonian or Lindblad operators.  Instead each atom interacts with and contributes to a local, mean field and sees the behaviour of other atoms through this mean field.  The mean field is a sum of the external incident field that excites the ensemble and a local field created by the driven and spontaneously emitting atoms (quantum emitters) in the ensemble:
\begin{equation}
\vec{E}(\vec{r},t) = \vec{E}_{inc}(\vec{r},t) +\vec{E}_{local}(\vec{r},t) .
\end{equation}
This method of using a mean-field interaction is used in numerous areas in computational physics, such as in polymer self-consistent field theory \cite{MscThesis}, and computational electrodynamics \cite{PhysRevLett.72.1651}.  For clarity, the ``mean'' in the mean field refers to a mean of the interactions between molecules and not a spatial mean of the fields themselves.  

This simplification allows the overall quantum state space to remain relatively small. For a system consisting of N four-level systems, the total directional state space ($M=4N$) is reduced to $4N$ quantum states and $3N$ local quantum interactions.  This greatly simplifies the equations, and allows us to solve the problem by evolving the density matrices locally with an efficient parallel implementation.  In this study, we model an ensemble of approximately 4000 atoms.  With the ensemble state basis reduced to a more manageable size, one now needs to determine how the quantum emitters create local fields. 

\subsection{Numerical Implementation}
\label{PSTD}
To implement this calculation numerically, we modify and extend the methodology used by Sukharev and Nitzan \cite{PhysRevA.84.043802}.  In our method, Maxwell's equations are solved numerically in time for a coarse-grained grid using a pseudo-spectral time domain method (PSTD) \cite{PTSD1, PTSD2}.   The choice of using a PSTD method over the FDTD method used in Ref.\cite{PhysRevA.84.043802} is largely because the PSTD method is computationally more efficient than the FDTD method \cite{PTSD1, PTSD2}.  There is also the added benefit of using a single lattice grid as opposed to the staggered grid required of the FDTD method \cite{kunz1993finite}.  A uniaxial perfectly matched layer (PML) \cite{PMLbook1,PMLbook2} is used to eliminate reflection at the boundaries, and to strongly attenuate the signal so as to prevent signal wraparound in the simulation \cite{PTSD1}.  For a plane wave, we modify the PML size and coefficients to reduce the relative reflected and wraparound field amplitudes to at most $10^{-5}$ of the incident field amplitude.

The simulation space is broken into a 3D computational grid, with each cell having associated with it an electric and magnetic field. This grid is chosen to be cubic with spacing of $l=1$ nm; this spacing corresponds to the interatomic spacing associated with the approximate atomic density used in the calculations ($N_A= 1 \times 10^{27}m^{-3}=l^{-3}$).  The individual quantum emitters are assumed to be point emitters.
The order of operations at each time-step is:
	\begin{itemize}
		\item The fields of the ``source cells" are updated analytically so that a plane wave is produced~\cite{PTSD1}.
		\item Maxwell's equations are solved numerically in time for this coarse-grained grid using the pseudo-spectral time domain method.  Firstly, the magnetic field $\vec{H}(\vec{r})$ is updated.
		\item If there is a quantum emitter present in a cell, the density matrix of that cell is evolved by solving the Lindblad-Von Neumann equation (\ref{eq:Lindblad}) using a fourth-order Runge-Kutta method\cite{press1989numerical}, and the electric fields at the previous time-step as input.  Going beyond previous studies \cite{PhysRevA.84.043802}, we include interatomic interactions in all three directions by implementing a generalized three-directional state basis described in subsection~\ref{3dbasis}.  
		\item The free current in each cell $\vec{J}(\vec{r})$ is determined for cells containing one or more quantum emitters.
		\item The free current is used to update the local electric field, $\vec{E}(\vec{r})$, using Maxwell's equations.
		\item The process is repeated and items of interest are recorded.
	\end{itemize}
Each simulation is run until the density matrix of the ensemble reaches an approximate steady state.  For a collection of ~4000 emitters, a simulation takes between 8-12 CPU days on 8 cores \cite{SHARCNET}.

\section{Example Application: Increasing Solar-Cell Efficiency} \label{sec:disorder}
Thermal upconversion is a very important process of interest in the design of highly efficient solar cells \cite{lenert2014nanophotonic}.  In silicon solar cells, electricity is only produced by photons with $\lambda < 1100 nm$ due to the band gap in silicon; therefore solar photons of much higher wavelengths are ``wasted" \cite{strumpel2007modifying}.  The goal of many in the scientific community is to design a nanoscale system that can blueshift significant amounts of infrared photons, thus recouping some of this under-utilized energy.

The Lorentz-Lorenz model of an atomic electron driven by an incident electromagnetic field predicts that the induced polarization has a frequency that is blueshifted~ \cite{lorentz_book}.  It can be expected that the induced electromagnetic field will also be at a blueshifted frequency compared to the incident field.  According to this model, a driven neodymium atom, for which the ground-to-excited-state transition energy is $\approx$ 1 eV, when placed onto silicon that has a bandgap of just above 1 eV, could theoretically blueshift the incident light, and increase the silicon's absorption.  We speculate that a dense arrangement of neodymium atoms on the silicon would be able to amplify this blueshifting effect.  Therefore we model a dense ensemble of atoms driven by a plane wave electromagnetic field, with an aim to exploit the macroscopic/collective effects amplified from the microscopic dynamics.

\begin{figure}
	\centering
	\includegraphics[width = 3in]{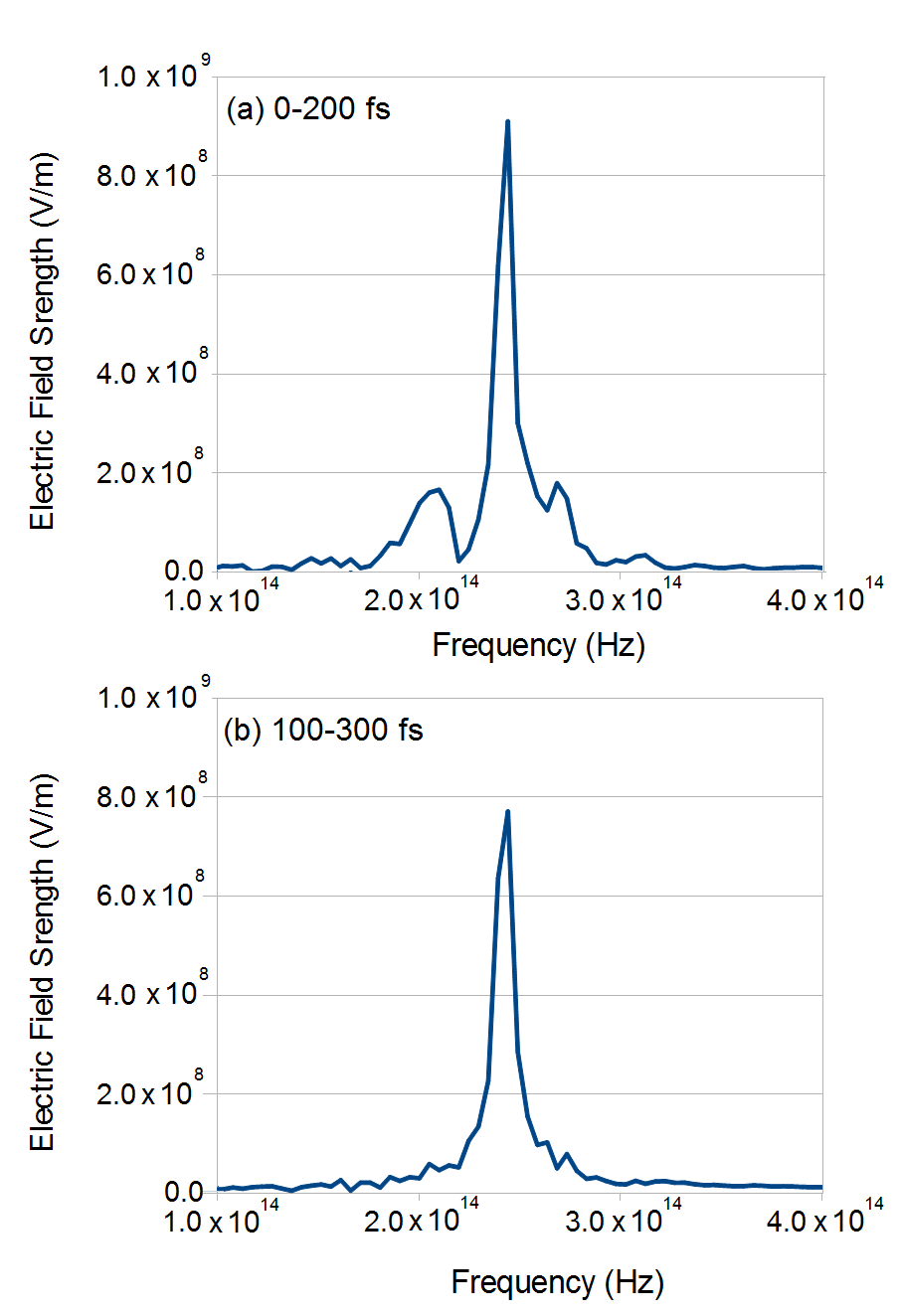}
	\caption{Fourier transform of the electric field over a 200fs time window at $\vec{r}$=(0, 13nm, 0), a point  outside a 10 nm radius spherical ensemble of atoms centered at the origin. Each atom in the ensemble has an energy level structure as shown in Fig.~\ref{fig:DirectionalStates}(b), with energy spacing between the ground and excited states of 1 eV, and spontaneous emission rates of 2.95MHz .   The number density of atoms in the ensemble is $4 \times 10^{27}$ atoms per cubic metre.  The incident plane wave electromagnetic wave of frequency 241 THz, and electric field amplitude 1.5GV/m is polarized in the $\hat{y}$-direction, and propagates along the $\hat{z}$ direction.   (a) Frequency components that appear in the time window 0 - 200fs after the start of excitation include a distinct blue-shifted peak.  (b) Frequency components that appear in the time window 100 - 300fs after the start of excitation. Notice that the blue-shifted frequency components have died out. \label{fig:shift_plasmon}}
\end{figure}

Using the methodology described in the previous section, we calculate the response of a dense quantum ensemble to a monochromatic, plane-wave, driving field of wavelength $197.5$ nm (corresponding to $1.0$ eV) in order to determine whether or not the frequency of the near field around the ensemble can be blueshifted.   The collection of dense quantum emitters is arranged in the form of a 10 nm nanosphere with an origin of coordinates at its centre.  The incident  monochromatic, plane wave is polarized in the $\hat{y}$-direction, and propagates along the $\hat{z}$-direction.   We monitor the electric field amplitude a short-distance (3 nm) outside the nanosphere for 200 fs (0 fs to 200 fs).   Taking a Fourier transform of this field amplitude, we see that the electromagnetic field around the nanosphere is no longer purely monochromatic (Fig \ref{fig:shift_plasmon}(a)) even if the input is.  There is a blueshifted component that appears.  Although this appears promising, if we continue the evolution and take a Fourier transform of the field for the window from 100-300 fs, the spectrum transforms to that depicted in  Figure \ref{fig:shift_plasmon}(b).  The blueshifted peak has disappeared.
This loss of upshifted frequencies at long times indicate that an ensemble of quantum emitters is not suitable for thermal upshifting in solar cells.  

\begin{figure*}
	\centering
	\includegraphics[width = 5in]{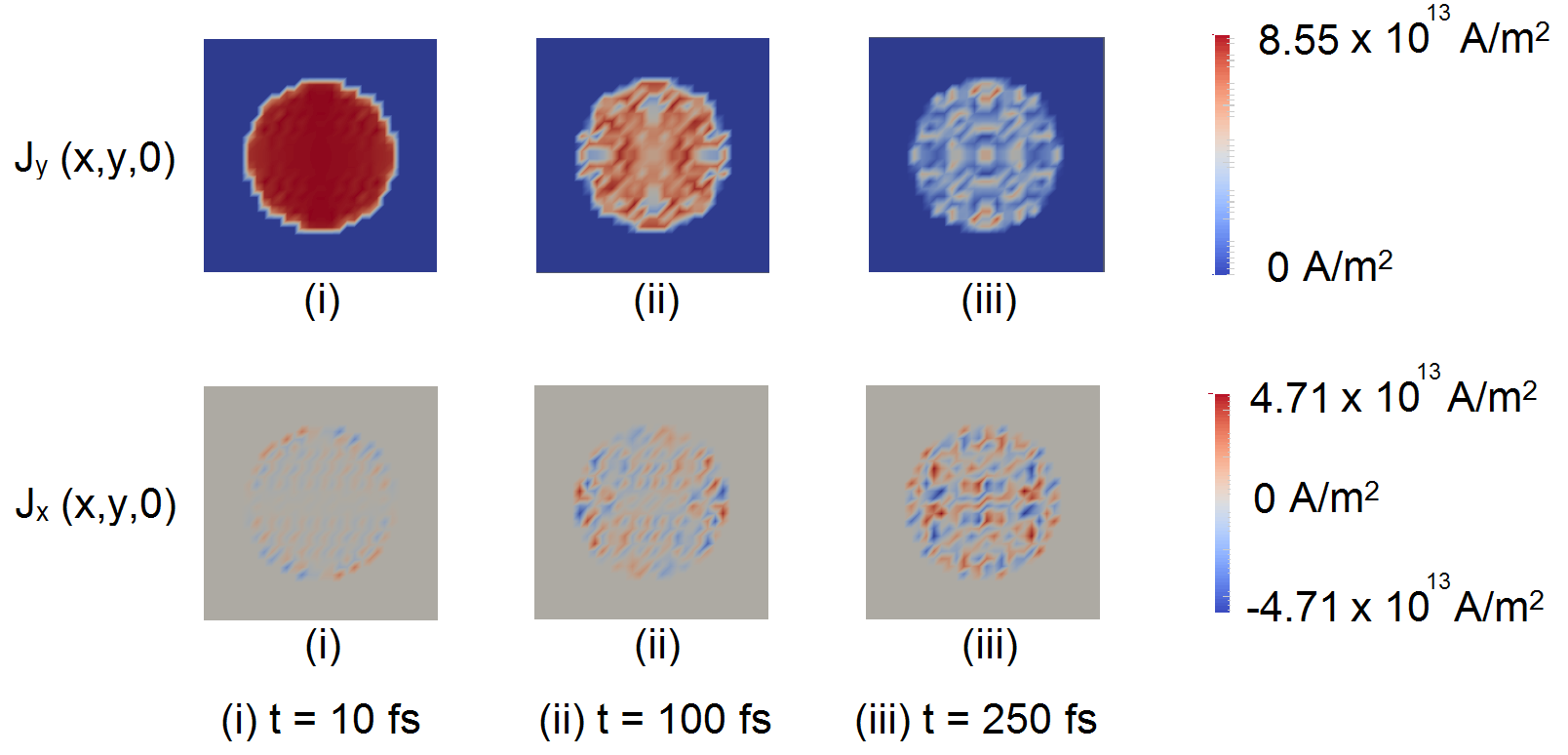}
	\caption{Snapshots of the spatial distribution of free-current density components $\vec{J}_y(\vec{r})$ (row (a)) and $\vec{J}_x(\vec{r})$($A/m^2$) (row (b)) in the $x-y$ plane (with $\hat{y}$ being horizontal) bisecting a 10 nm nanosphere of atoms at times (i) 10 fs, (ii) 100 fs and (iii) 250 fs after start of excitation. Parameters of the ensemble and the incident field are the same as in Fig.\ref{fig:shift_plasmon}.  The snapshots show that at early times, there are ordered patterns in the spatial distribution of the free current density; and as time goes on, disorder sets in due to interatomic interactions, finally ending in a disordered `phase'.    \label{fig:order_disorder}}
		
\end{figure*}

In order to probe why the frequency-shifted components disappear, we examine the spatial distribution of free-current density components ($J_\eta (\vec{r})$) in the nanospherical ensemble as a function of time.  Snapshots of the free-current in the $xy$-plane are depicted in Fig \ref{fig:order_disorder}. It is immediately seen that the distribution of free currents in the ensemble becomes disordered as time goes on.  Initially, the ensemble responds to the incident field in what is effectively an ordered phase; all the individual atoms respond to the field by oscillating in an identical manner. This phase is characterized by a near-uniform free current distribution anti-aligned with the incident field polarization.  The spatial distribution of the free currents in directions perpendicular to the incident field polarization show weak, quadrupolar patterns.  At later times, due to the build-up of electric field components perpendicular to the incident polarization, the overall ordered pattern is lost, and small instantaneous domains are formed that do not move in phase with one another. These two phases that we refer to as `ordered' and `disordered' correspond to the two time windows; one that has a blueshifted frequency and one that does not.  The time-scale of this onset  of disorder ($\approx 28 fs$) in the free-current distribution is much faster than what one would expect from the normal spontaneous emission rates of the individual emitters ($1/\gamma_0 \approx 344 ns$).

\begin{figure}
	\centering	
	\includegraphics[width = 3in]{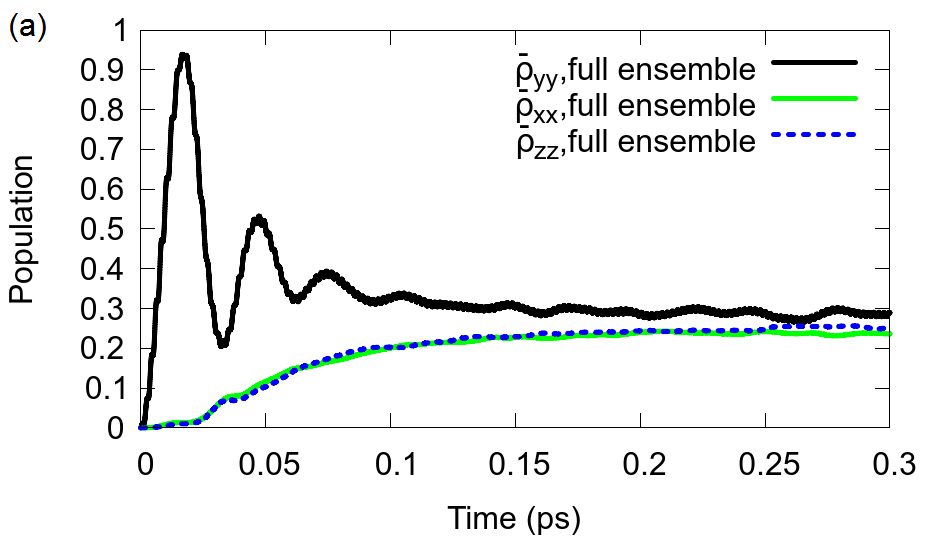}	
	\includegraphics[width = 3in]{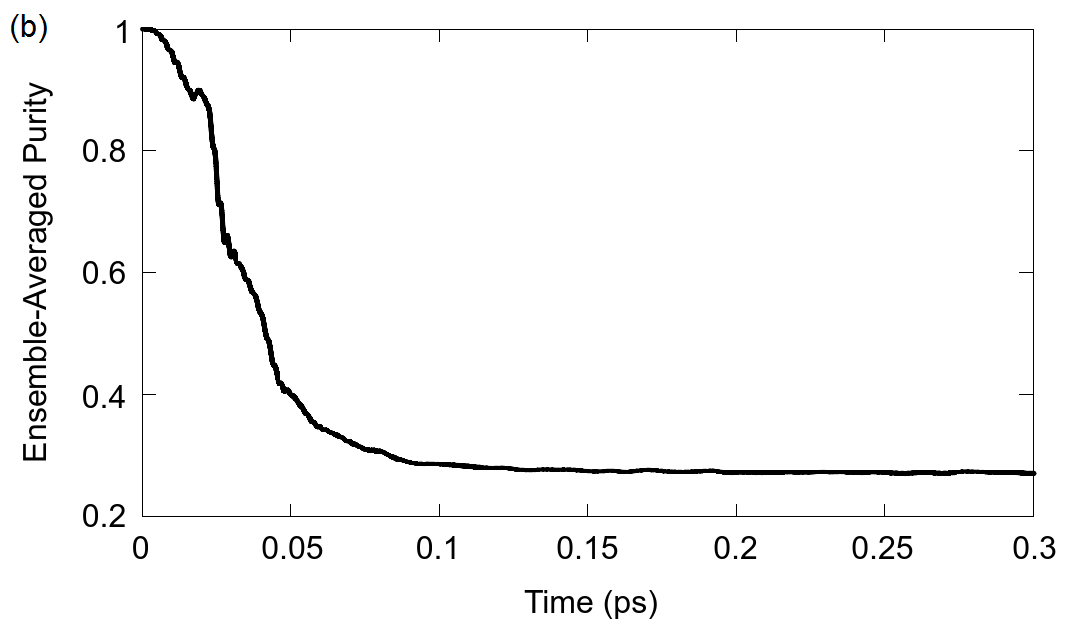}
	\caption{(a) Spatially-averaged populations in the $\hat{x}$, $\hat{y}$ and $\hat{z}$-directional excited states and (b) ensemble-averaged purity for a 10 nm radius nanosphere of atoms with atomic number density $N_A = 4.0 \times 10^{27} m^{-3}$. Other parameters of the ensemble and the incident field are the same as in Fig.\ref{fig:shift_plasmon}.  \label{ensemble-pop}}.
\end{figure}

Examination of the spatially-averaged ensemble density matrix ($\bar{\rho} = \frac{1}{V}\int d^3 \vec{r}\rho(\vec{r}) =\frac{1}{N} \sum^N_n \rho_n$) reveals some interesting connections between the macroscopic and microscopic dynamics.  In Fig.~\ref{ensemble-pop}, we see that the ensemble-averaged excited-state population that lies along the incident polarization axis ($\bar{\rho}_{yy}$) appears to quickly reach a steady-state.  As the free-current distribution quickly becomes disordered, non-directly-driven excited states ($\ket{e_x}$ and $\ket{e_z}$) gain and retain state population, as seen from the increase in $\bar{\rho}_{xx}$ and $\bar{\rho}_{zz}$.  This directly shows that inter-atomic interactions (mediated through a mean field) with strong driving fields lead to a mixing of multi-directional excited states.  As all of the ensemble state populations rapidly reach an approximate steady-state that oscillates only with the incident frequency, the time-averaged coherences in the rotating frame reduce to a small net coherence oscillating in the incident field polarization direction with the frequency of the incident field.  By examining the purity ($Tr(\bar{\rho}^2)$) of the ensemble in Fig.\ref{ensemble-pop}(b), it is seen that the ensemble state undergoes decoherence over the same timescale as the population leakage.  

The onset of disorder in the current density distribution is directly linked to the fact that population from the excited state corresponding to the polarization direction of the incident light (in our case $|e_y\rangle$) is redistributed due to interatomic interactions into the other excited states ($|e_x\rangle$ and $|e_z\rangle$), which in turn is linked to decoherence in the ensemble state.  This ``directional state leakage", and associated decoherence effects occur on a time-scale that is much faster that what would be predicted by normal spontaneous emission by several orders of magnitude (the lifetime of the ensemble excited state about $28$ fs  in comparison to the lifetime of a single 2LS which is $\approx 344 ns$).   Note that this disorder is purely an ensemble effect; the local purity of individual coarse grains remains close to unity on this time-scale since the individual spontaneous emission rate is low ($2.95 \times 10^6$ Hz).

By examining the dynamics of the strongly-driven, dense ensemble of 2LS, we find that interatomic interactions create strong disorder in the ensemble states over a characteristic time.  This disorder imposes an overall effective lifetime for quantum scattering effects.  

\begin{figure}
	\centering
     \includegraphics[width = 3in]{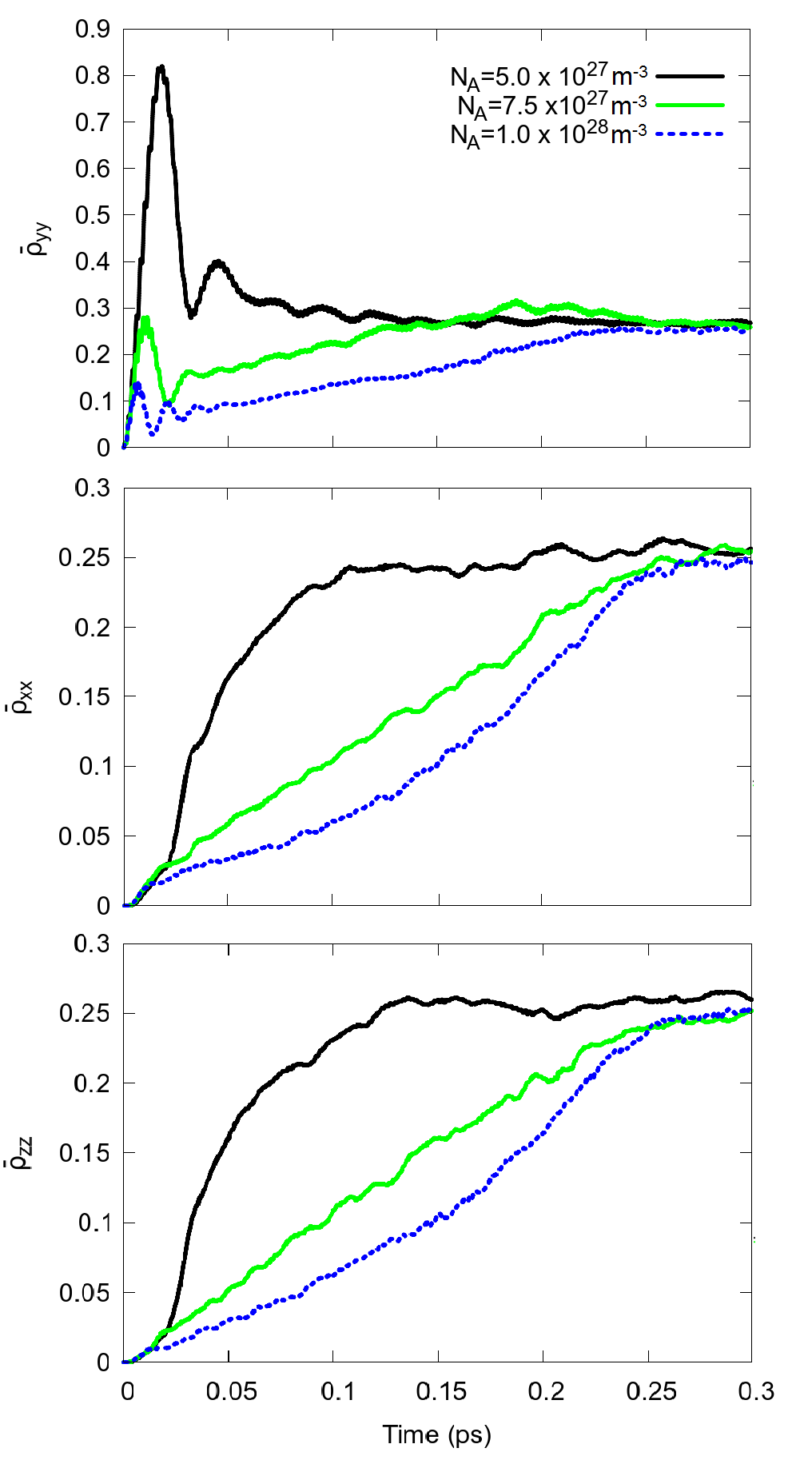}
	\caption{Spatially averaged populations ($\bar{\rho}_{xx}$, $\bar{\rho}_{yy}$, $\bar{\rho}_{zz}$) in the $\hat{x}$, $\hat{y}$  and $\hat{z}$-directional excited states for a 10 nm radius spherical ensemble of atoms with varying number densities.  Parameters of the incident field are the same as in Fig.\ref{fig:shift_plasmon}.   \label{fig:dens_vary_pop_y}}
\end{figure}

Figure \ref{fig:dens_vary_pop_y} shows the ensemble-averaged excited state populations as a function of increasing number density.  At very low number density, the ensemble-averaged excited state population oscillates much in the same way as a single, driven two-level system with spontaneous emission.  Since the interatomic interactions are low, the population in the non-directly driven excited states isn't much.  As the number density of atoms increases, the interatomic interactions cause population leakage into the non-directly driven excited states.  At the same time, we see that the oscillation in the directly-driven excited state is damped much more quickly than the low density case.  Increasing interatomic interactions appear to increase the rate of spontaneous emission in the ensemble, which we already saw is linked to the onset of disorder in the free-current density.  As the number density increases further, screening makes it more difficult to excite population into the directly-driven excited state, and hence the populations in the non-directly driven excited states increases at slower rates.

The presence of these extremely strong, decoherent processes in a driven quantum ensemble has immediate consequences for the numerical modelling of a driven ensemble of quantum emitters. Firstly, these results indicate that the ``short-pulse method''  \cite{PhysRevA.84.043802}(the use of ultra-short, sub-fs pulses to determine continuous scattering amplitudes) may not be generally applicable when modelling quantum systems driven at high intensities.   Secondly, these results indicate that for an ensemble of quantum emitters, a one or two-directional basis set (such as in Ref. \cite{rand2008,DIET,PhysRevA.84.043802})is insufficient to fully capture inter-atomic interactions, and can lead to overestimates in their long-term coherent behaviours at high densities.  This indicates that for a general ensemble of dense emitters, a full directional state basis is required.  Our calculation therefore goes beyond the standard approximations by using a plane wave excitation, and a full three-dimensional state basis.

\subsection{Quantifying Disorder in Driven, Dense Quantum Ensembles}

The ensemble-averaged excited state density in the incident field polarization direction $\bar{\rho}_{yy}$ can be fit to a phenomenological model of a driven two-level system in which there is spontaneous decay from the excited state to the ground state, as well as a loss of population density.
\begin{equation}
 \bar{\rho}_{yy} = a \exp(-\gamma_{ens} t) \cos(\Omega t) + b + c \exp(-g t),
 \end{equation}
 \noindent where a, b, and c are dimensionless constants, $\gamma_{ens}$ is analogous to the damping rate of the driven excited state ($\hat{y}$) that we call ``the disorder-onset rate",  $g$ represents the rate at which state population ``leaks'' from the $\ket{e_y}$ state to $\ket{e_x}$ and $\ket{e_z}$ excited states, and $\Omega$ is the Rabi frequency that is proportional to the electric field amplitude of the near-resonance driving field.

Just as the spontaneous emission rate of an individual quantum state tells us how long a 2LS can remain viable as a qubit, the effective disorder-onset rate of the system tells us how long true quantum behaviour stays relevant in the ensemble.  


A table summarizing the fits for disorder-onset rates and state leakage rates as a function of increasing number density of the atoms in the ensemble can be found in Table \ref{table:fit1}.
\begin{table}
	\centering
	\begin{tabular}
		{ | c | c | c |}
		\hline
		Number Density $N_A$ ($m^{-3}$) & $\gamma_{ens}$ (Hz) & $g$ (Hz) \\ \hline \hline
		
		$1 \times 10^{27}$ & $6.243 \times 10^{11} $  & $8.983 \times 10^{11}$ 
		 \\ \hline 
		 		$2.5 \times 10^{27}$ & $1.455 \times 10^{13}  $ & $6.173 \times 10^{12}$ 
		 		\\ \hline 
		 				$4 \times 10^{27}$ & $3.555 \times 10^{13} $ & $1.845 \times 10^{13}$ 
		 				\\ \hline 
		 						$5 \times 10^{27}$ & $5.072 \times 10^{13}$  & $2.637 \times 10^{13}$ 
		 						\\ \hline 
		 								$7.5 \times 10^{27}$ & $5.305 \times 10^{13} $ & $9.193 \times 10^{12}$ 
		 								\\ \hline 
		 										$1 \times 10^{28}$ & $5.194 \times 10^{13} $ &  $1.475 \times 10^{12}$ 
		 										\\ \hline 

	\end{tabular}
		\caption{Disorder-onset rates ($\gamma_{ens}$), and excited-state population leakage rate ($g$) for a 10 nm radius spherical ensemble of atoms with varying number density ($N_a$).  The amplitude of the driving electromagnetic wave is E=$1.5 \times 10^9$ V/m. The spontaneous emission rate of a single atom in the ensemble is 2.95MHz.\label{table:fit1}}
\end{table}

\begin{figure}
	\centering
	\includegraphics[width = 3in]{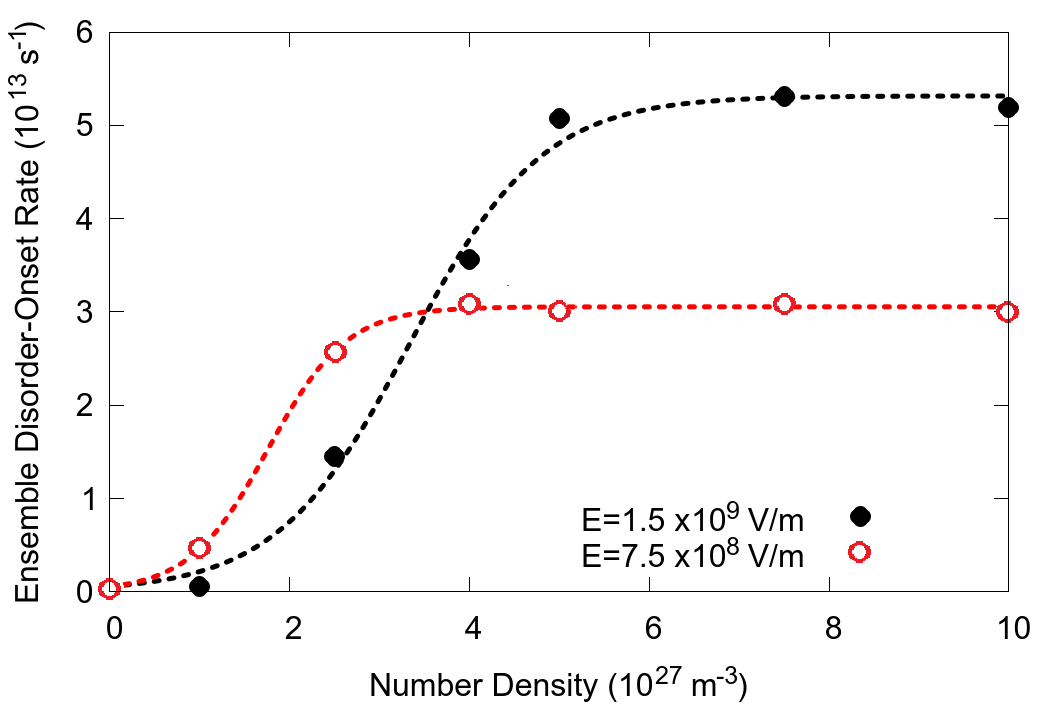}
		\begin{tabular}
		{ | c | c | c | c |}
		\hline
		Intensity ($V/m$) & $L$ (Hz) & $a$ ($m^-3$) & $k$ ($m^3$) \\ \hline \hline
		$1.5 \times 10^9$ & $5.316 \times 10^{13} $  & $3.337 \times 10^{27}$ & $1.353 \times 10^{-27}$
		\\ \hline 
		$7.5 \times 10^8$ &  $3.055 \times 10^{13}$  & $1.762 \times 10^{27}$ & $2.286 \times 10^{-27}$
		\\ \hline 	
	\end{tabular}
	\caption{Effective ensemble disorder-onset rates ($\gamma_{ens}$) for a 10 nm radius spherical ensemble of atoms with varying number density ($N_a$), for two different amplitudes of the driving plane-wave electromagnetic wave.  The spontaneous emission rate of a single atom in the ensemble is 2.95MHz. The data are fit to a logistic function as in Eq.~\ref{eq:logistic}.  The saturation value of the disorder-onset rate $L$ increases as the amplitude of the driving field increases. \label{fig:logistic_fit}}
\end{figure}

As the number density increases, the disorder onset rate $\gamma_{ens}$ increases.  At very high number density, $\gamma_{ens}$ becomes so large that the $\ket{e_y}$ state cannot be significantly populated, so the ``leakage'' to other directional states starts to disappear.  We note that the onset of disorder in denser ensembles is largely dominated by $\gamma_{ens}$.  The dependence of $\gamma_{ens}$ as a function of number density ($N_a$) is plotted in Figure \ref{fig:logistic_fit} for a dense ensemble driven with strong fields ($\Omega>>\gamma_0$).  From this figure, it is clear that a strongly driven, dense quantum ensemble experiences a fast (compared to a single atom's spontaneous emission rate $\gamma_0 = 2.95 \times 10^6$ Hz) onset of disorder, and the disorder-onset rate increases as the density of atoms in the ensemble increases.  This indicates that both rapid onset of disorder and leakage to three-directional states via interatomic interactions are important in the dynamics of a strong driven ensemble of atoms.  Any quantum control calculations that are applied to dense collections of atoms should not use short pulse methods and/or reduced basis sets that ignore directional states unless they are driven by extremely rapid pulses or have a low number density.

The dependence of $\gamma_{ens}$ on $N_A$ is particularly interesting.  Figure~\ref{fig:logistic_fit} shows that this dependence is nonlinear and although it increases at low densities, the disorder-onset rate ($\gamma_{ens}$) slows down at high densities and converges to a saturation value.  This behaviour appears to be best described by a saturation curve that takes the form of the logistic function \cite{gershenfeld1999nature} 
\begin{equation}
\gamma_{ens} = \frac{L}{1+\exp(-k(x-a))}, \label{eq:logistic}
\end{equation}
\noindent where, $L$ is the saturation value of $\gamma_{ens}$, $k$ is a rate constant, $x$ is the number density and $a$ is the inflection point of the number density at which the disorder onset rate begins to saturate.  This saturation curve is typically used in evolutionary systems in which there is a competition between different processes \cite{verhulst1845mathematical}.  In this particular ensemble system, there is a competition between the incident field that is trying to force the ensemble to oscillate coherently, and the disorder (i.e. the mean-field mediated inter-atomic interactions) that is trying to prevent this coherent oscillation.

Figure \ref{fig:logistic_fit} shows the fit of the disorder onset rate $\gamma_{ens}$ to the logistic function (Eq.~\ref{eq:logistic}) for two different incident field intensities.  One conclusion that can be easily drawn from such fits is that, as the intensity of the incident light is increased, the saturation point ($L$) of the disorder-onset rate also increases.  This is because, at higher intensities, the coherent driving by the incident field excitation is able to more strongly overcome the decoherence caused by interatomic interactions.

This dependence of the disorder-onset rate ($\gamma_{ens}$) on the amplitude of the driving indicates that for strongly-driven, dense quantum systems, {\emph{the disorder-onset rate} is dependent on the density matrix, and therefore \emph{is time-dependent}.  For dense collections of quantum emitters, a better model of $\gamma_{ens}$ than a constant value, would be to estimate it by using the quantum state of the ensemble.  

\section{Modelling Dense Ensemble Dynamics with Single Particle Techniques}\label{sec:single}

Examining the dynamics of a driven, nanoscale ensemble of quantum systems, one notable observation is that the evolution of the ensemble state population in the incident field polarization direction is qualitatively similar to that of a driven two-level system with two competing decoherence mechanisms --- spontaneous emission, and a loss of population from the excited state parallel to the incident field polarization.  Therefore we aim to approximate this behaviour with a single-particle model by modifying the decoherence scheme.  

This single-particle model should be similar in nature to the individual particles that make up the ensemble.  Its basis consists a ground state $\ket{g}$ and three directional excited states, $\ket{e_x}$, $\ket{e_y}$ and $\ket{e_z}$ and it is excited by an incident plane wave. The Hamiltonian of this system, after making the rotating wave approximation is
\begin{equation}		
H =  \left ( \begin{matrix}  
0 & \frac{\hbar\Omega_{x}}{2} & \frac{\hbar\Omega_{y}}{2} & \frac{\hbar\Omega_{z}}{2} \\
\frac{\hbar\Omega_{x}^*}{2}&-\bigtriangleup&0&0\\
\frac{\hbar\Omega_{y}^*}{2} &0&-\bigtriangleup&0\\
\frac{\hbar\Omega_{z}^*}{2} &0&0&-\bigtriangleup\\
\end{matrix} \right ),
\end{equation}
\noindent where, $\bigtriangleup$ represents the detuning between the atomic transition frequency and the frequency of the incident light, and the Rabi frequency-like terms $\Omega_i, i=x,y,z$ are proportional to the electric field amplitudes in each of the three Cartesian directions.

For this Hamiltonian, the electric field terms included are the external incident field ($E_y$), and perpendicular scattered field components $E_x$ and $E_y$ that are much smaller than the incident field.  For the perpendicular scattered field components, we assume they arise from the field of a dipole with $E_{x,z} \approx E_y \frac{\mu}{e r}sin(\theta) \hat{\theta}$ \cite{sharma}.  In this case, $r= \sqrt[3]{\frac{3 \sqrt{8}}{4 N_a  \pi}}$ is the separation between diagonal nearest neighbours, $\theta = \pi/4$ is the angle between them, $e$ is the charge of an electron and $\mu$ is the transition dipole moment.  For dense systems, the magnitude of this scattered field is about 1-2 orders of magnitude less than that of the incident field.  

In the ensemble, an individual quantum system can spontaneously emit radiation from the $\ket{e_x}$, $\ket{e_y}$, and $\ket{e_z}$ excited directional states with rates $\gamma{xg}$, $\gamma{yg}$ and $\gamma{zg}$ respectively.  This emitted radiation can then excite either the $\ket{g} \rightarrow \ket{e_x}$, $\ket{g} \rightarrow \ket{e_y}$, or $\ket{g} \rightarrow \ket{e_z}$ transitions in nearby atoms.  This process is similar to the Forster-Resonance Energy Transfer (FRET) process commonly seen in biophysical systems \cite{nanopticsbook}.  We adopt this FRET model to the decoherence in our single particle model.  Decoherence couplings are added that look like forbidden electric-dipole transitions as shown in Figure \ref{fig:DecayCoupledDirectionalStates}.  Although these transitions look similar to spontaneous emission, they do not result in net emission of a photon.  Rather, they represent the emission of a photon and the reabsorption of that photon by another transition in an adjacent atom.  This makes these transition rates behave more like dephasing rates ($\delta_{ij}$), as they do not emit energy from the system.  $\delta_{ij}$ represents the dephasing rate due to emission of a photon from the state $\ket{e_i}$ of one atom that is absorbed by a neighbouring atom that is excited to state $\ket{e_j}$.  $\delta_{xx}$, $\delta_{yy}$ and $\delta_{zz}$ are referred to as ``parallel" dephasing rates, whereas $\delta_{xy}$, $\delta_{yz}$, and $\delta_{zx}$ are referred to as ``perpendicular" dephasing rates.   A diagram of all the decoherence processes in the two-level, directional state basis of the single-atom model is provided below in Figure \ref{fig:DecayCoupledDirectionalStates}.

\begin{figure}
	\centering
	\includegraphics[width = 3in]{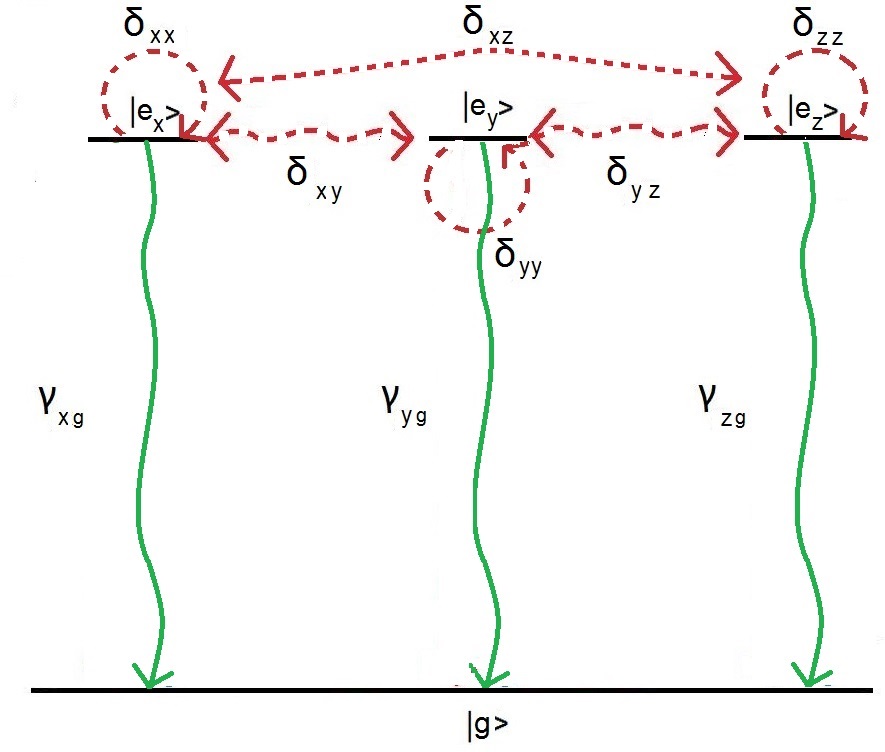}
	\caption{ Modified decoherence structure in the single-particle model of a driven atomic ensemble.  When significant inter-atomic interactions are present in an ensemble, it becomes possible for the spontaneously emitted radiation from the excited state of one atom to excite state population from the ground state of a nearby atom.  This emission followed by absorption is modelled by a dephasing process between states that have electric-dipole forbidden transitions ($\delta_{ij}$'s in red). These dephasing rates do not affect the total state population, they only reduce the overall coherence of the single particle state that models the ensemble. 	\label{fig:DecayCoupledDirectionalStates}}
\end{figure}

\subsection{Estimating Decoherence Rates}
We want to estimate the decoherence rates $\gamma_{ig}$'s and $\delta_{ij}$'s that will be inputs into the single particle model.  Let $\vec{E}_d$ be the amplitude of the field driving the ensemble, and $\vec{E}_{local}$ be the local field at the location of the atom.  Let $\gamma_0$ be the vacuum spontaneous emission rate of a single two-level atom.

To calculate the spontaneous emission rate $\gamma_d$ from an excited state to a ground state of an atom in an ensemble, one can define an `enhancement factor' by comparing the power emitted by an atom in an ensemble $P$ to that which it emits in free space $P_0$, calculated via the Larmor formula.  This takes the form:
\begin{equation}\begin{aligned}
\frac{\gamma_d}{\gamma_{0}} &= \frac{P}{P_{0}} =   \frac{Re(\vec{j}_d^* \cdot \vec{E}_{local})}{Re(\vec{j}_d^* \cdot \vec{E}_{d})},
\end{aligned}
\end{equation}
\noindent where $\vec{j}_d$ is the free current of the transition.  The local field $\vec{E}_{local}$ is the sum of the driving field $\vec{E}_d$, and the field scattered by other atoms $\vec{E}_{ext}$.  Therefore,
\begin{equation}\begin{aligned}
\frac{\gamma_d}{\gamma_{0}} &= \frac{P}{P_{0}} =  1 +  \frac{Re (\vec{j}_d^* \cdot \vec{E}_{ext})}{Re(\vec{j}_d^* \cdot \vec{E}_{d})}.
\label{eqn:hi_td_decay_enhance},
\end{aligned}
\end{equation}

If the ensemble contains many strongly interacting quantum elements, the decay rate enhancement in various directions will be a complicated function of time and therefore cannot be easily evaluated with a single, constant, enhancement.  If the transitions are oscillating dipole emitters; however, Eq.~\ref{eqn:hi_td_decay_enhance} can be simplified to \cite{nanopticsbook},
\begin{equation} \begin{aligned}
\frac{\gamma_d}{\gamma_{0}} = \frac{P}{P_{0}} &= 1 - \frac{6 \pi \epsilon_0}{|\mu_0|^2} \frac{c^3}{\omega^4} Re(\vec{j_0}^* \cdot \vec{E_{ext}}), \\
&=  1 + \frac{6 \pi \epsilon_0}{|\mu_0|^2} \frac{1}{k^3} Im(\vec{\mu_{0}}^* \cdot \vec{E_{ext}}).	
\label{eqn:td_decay_enhance}
\end{aligned}\end{equation}
In the single particle model, there are no fields due to scattering from other atoms, i.e., $E_{ext}=0$.  Therefore the spontaneous emission rates $\gamma_{ig}$ are all equal to $\gamma_0$.

The dephasing rates ($\delta_{ij}$) associated with energy transfer between atomic transitions can be calculated by the following process \cite{nanopticsbook}.  The magnitudes of the dephasing rates depend on the excitation transfer between atoms.   At different spatial locations, these dephasing rates can be quantified by
\begin{equation}
\frac{\delta_{i \rightarrow j}}{\gamma_0} = \frac{P_{i \rightarrow j}}{P_0},
\end{equation} 
\noindent where $\delta_{i \rightarrow j}$ is the rate of energy transfer from transition $i$ in one atom ($\ket{e_i} \rightarrow \ket{g}$) to transition $j$ ($\ket{g} \rightarrow \ket{e_j}$) in a neighbouring atom, and $P_{i \rightarrow j}$ is the power received by the ``acceptor'' transition ($j$) from the field created by the ``donor'' transition ($i$).  $P_{i \rightarrow j}$ is computed by
\begin{equation}
P_{i \rightarrow j} = \frac{1}{2} Re(\vec{j}^*_j(\vec{r_j}) \cdot \vec{E}_i(\vec{r_i})),
\label{eqn:Fret_Power2}
\end{equation}
\noindent where $\vec{j}^*_j(\vec{r_j})$ is the free current of the acceptor and $\vec{E}_i(\vec{r_i})$ is the field created by the donor. 


Starting with the near field of an radiating point dipole
\begin{equation}
\vec{E_i}(\vec{r}) = \frac{1}{4\pi\epsilon_0} \left( \frac{3(\vec{\mu_i}\cdot \hat{r})\hat{r}-\vec{\mu_i}}{r^3} \right ).
\end{equation}
where $\vec{E_i}(\vec{r}) $ is the electric field, $\vec{\mu_i}$ is the dipole moment of transition $i$ in a single particle and $\vec{r}$ is the spatial position, we can assume that the atoms are spherically distributed two atomic radii apart ($1/r^3 =\frac{1}{8} \frac{4 \pi}{3} N_a$).  This yields
\begin{eqnarray}
\vec{E_i}(\vec{r}) &=& \frac{1}{4\pi\epsilon_0} \frac{1}{8} \frac{4 \pi}{3} N_a \left(3(\vec{\mu_i}\cdot \hat{r})\hat{r}-\vec{\mu_i} \right ) \nonumber\\&=& \frac{N_a}{24 \epsilon_0} |\mu_i|\left( 3(\hat{\mu_i}\cdot \hat{r})\hat{r}-\hat{\mu_i} \right ).
\end{eqnarray}
Therefore the power transferred due to interaction between two individual particle transitions ($i$ and $j$), assuming that $\vec{j} \sim \omega \vec{\mu}$
\begin{equation}
P_{i \rightarrow j} = \frac{1}{2} Re(\vec{j}^*_j(\vec{r}) \cdot \vec{E}_i(\vec{r})) \approx \frac{1}{2} \omega\vec{\mu_j} \cdot \vec{E}_i(\vec{r})
\end{equation}
becomes
\begin{equation}
P_{i \rightarrow j} = \frac{1}{2} \omega\vec{\mu_j} \cdot (\frac{N_a}{24 \epsilon_0} |\mu_i|\left( 3(\hat{\mu_i}\cdot \hat{r})\hat{r}-\hat{\mu_i} \right )).
\end{equation}
Given that the dipole moments for each transition are the same ($|\mu_{i,j}|=|\mu|$) (as all atoms are identical), this reduces to
\begin{eqnarray}
P_{i \rightarrow j} &=& \frac{N_a \omega}{48 \epsilon_0} |\mu|^2\left (\hat{\mu_j} \cdot\left( 3(\hat{\mu_i}\cdot \hat{r})\hat{r}-\hat{\mu_i} \right )\right ) \\ 
&=& \frac{N_a \omega}{48 \epsilon_0}  |\mu|^2\left ( 3(\hat{\mu_i}\cdot \hat{r})(\hat{\mu_j} \cdot\hat{r})-(\hat{\mu_j} \cdot\hat{\mu_i}) \right ).
\end{eqnarray}
With this, we can calculate $\delta_{i \rightarrow j}$ by normalizing to the power output of a classical oscillating dipole
\begin{equation}
\frac{\delta_{i \rightarrow j}}{\gamma_0} = \frac{P_{i \rightarrow j}}{P_0} = \frac{\frac{N_a \omega}{48 \epsilon_0}  |\mu|^2\left (3(\hat{\mu_i}\cdot \hat{r})(\hat{\mu_j} \cdot\hat{r})-(\hat{\mu_j} \cdot\hat{\mu_i})\right )}{\frac{\mu_0 \omega^4 |\mu|^2}{12\pi c}}.
\end{equation}
This yields
\begin{eqnarray}
\frac{\delta_{i \rightarrow j}}{ {\gamma_0}} &=&\frac{ N_a \pi c^3} {4 \omega^3}    \left( 3(\hat{\mu_i}\cdot \hat{r})(\hat{\mu_j}\cdot\hat{r})-\hat{\mu_j}\cdot\hat{\mu_i} \right ).
\end{eqnarray}
Lastly we add a factor of $\sqrt{\rho_{ii} \rho_{gg}} \sqrt{\rho_{jj} \rho_{gg}}$ which serves as an estimate of the fraction of atoms in the ensemble that experience the $\ket{i} \rightarrow \ket{j}$ energy transfer.  
\begin{eqnarray}
\frac{\delta_{i \rightarrow j}}{ {\gamma_0}} &=&\frac{ N_a \pi c^3} {4 \omega^3}    \left( 3(\hat{\mu_i}\cdot \hat{r})(\hat{\mu_j}\cdot\hat{r})-\hat{\mu_j}\cdot\hat{\mu_i} \right )\nonumber \\ && (\sqrt{\rho_{ii} \rho_{gg}} \sqrt{\rho_{jj} \rho_{gg}})\label{eqn:para};
\end{eqnarray}
For the ``parallel'' transitions (for example $\delta_{xx}$), we use Equation \ref{eqn:para} and normalize to the power of a radiating dipole of the transition frequency $\omega$,
\begin{equation}
\frac{\delta_{i \rightarrow j}}{ {\gamma_0}} =\frac{ N_a \pi c^3} {2 \omega^3} (\sqrt{\rho_{ii} \rho_{gg}} \sqrt{\rho_{jj} \rho_{gg}}).
\end{equation}

For the transitions that are ``perpendicular'' (for example $\delta_{xy}$), we use the nearest diagonal neighbour, instead of the nearest neighbour, as this diagonal neighbour is the closest lattice site in which a dipole can produce radiated fields in a perpendicular direction to its dipole moment.  This involves dividing Equation \ref{eqn:para} by $\frac{1}{\sqrt{8}}$ since $r' = \sqrt{2} r$ and therefore $\theta = \pi/4$.  The dephasing rate of a perpendicular transition is calculated as,
\begin{equation}
\frac{\delta_{i \rightarrow j}}{ {\gamma_0}} =\frac{ 3 N_a \pi c^3} {16 \sqrt{2} \omega^3} (\sqrt{\rho_{ii} \rho_{gg}} \sqrt{\rho_{jj} \rho_{gg}}).
\end{equation}

Placing these decoherence parameters into the single-particle Liouville equation, and solving numerically, yields excited state populations depicted in Figures \ref{fig:hola} (a) and (b).  The single-particle state calculation is overlaid with the ensemble-averaged calculation described in the previous section.  Comparing the results of the single particle approximation to the full ensemble calculation, we see that there is relatively good agreement between the two methods.  The two curves are not identical, however they are close enough to suggest that this single particle, modified-decoherence scheme captures a significant amount of the underlying physical processes involved, and that the FRET process is a good model of interatomic interactions in our mean-field calculation.

\begin{figure}
	\centering
	\includegraphics[width = 3in]{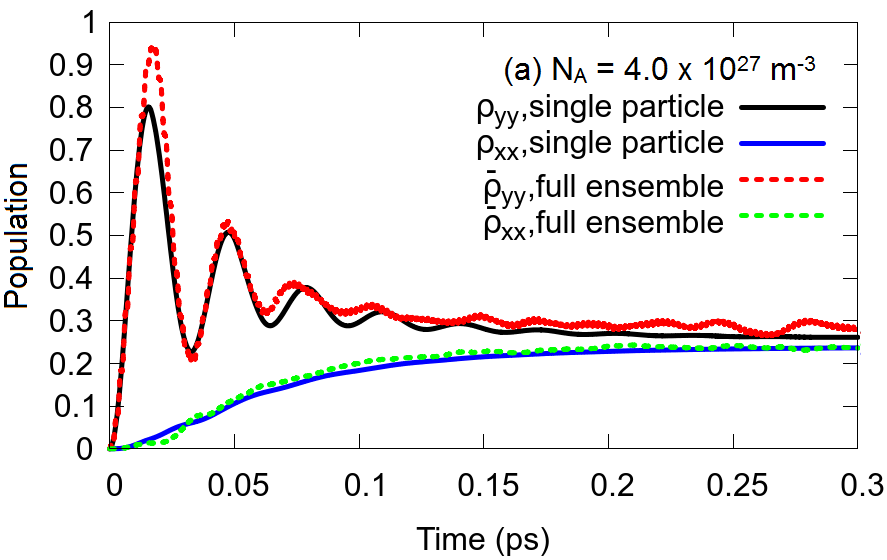}
	\includegraphics[width = 3in]{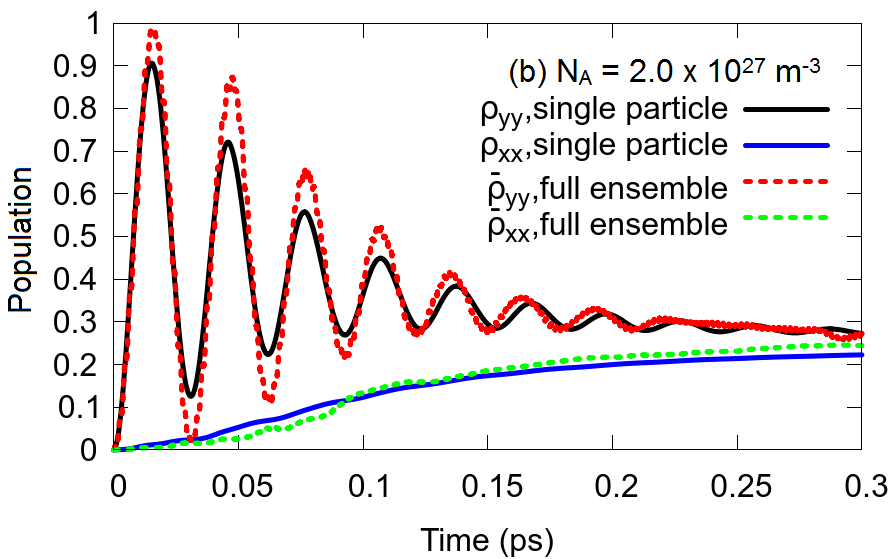}
	\caption{Comparison between single particle model calculation, and the mean field PSTD calculation of  spatially averaged excited state populations for a 10 nm radius spherical ensemble of atoms.  The amplitude of the driving electromagnetic wave is E=$1.5 \times 10^9$ V/m. The number density of atoms in the ensemble are (a) $4.0 \times 10^{27}$ and (b) $2.5 \times 10^{27}$ atoms per cubic metre.   \label{fig:hola}}
\end{figure}

The success of the effective single particle model shows that a FRET-like decoherence process takes place in a dense, driven ensemble.  The full calculation required $\approx 16$ CPU days of runtime; in comparison the single particle calculation required $\approx 2$ CPU minutes of runtime.  Thus, the single particle model can provide a reasonably accurate, quick estimate of the quantum dynamics in an ensemble before attempting a full calculation.

The main limitation of this single-particle model is that it does not explicitly include coherent scattering of a field emitted by one emitter from another emitter.  That is, in the Hamiltonian, only the incident electromagnetic field appears.  In reality, this Hamiltonian should also depend on the instantaneous state and overall geometry of the ensemble. Another limitation of this model it that it assumes that only the single, nearest neighbour interactions are relevant to the couplings; in truth, farther couplings and interference effects between atoms are required to increase the model's accuracy.  In future work, one could improve this model by adopting a more robust coupling geometry to account for scattered driving fields, and farther neighbours.  

		\section{Conclusion} \label{sec:conclusion}

We have studied the behaviour of a dense ensemble of quantum emitters driven by an intense, electromagnetic plane wave.  The state of each quantum emitter evolves according to the Lindblad-Von Neumann equation.  The evolution of the ensemble reflects not only the interaction between the driving field and individual atoms, but also the interactions between individual emitters.  To study this evolution, we have implemented a coarse-grained, mean field method based on the PSTD technique in which the Lindblad-Von Neumann equation for each quantum emitter is solved in conjunction with a solution to Maxwell's equations over the whole ensemble.  In order to correctly model the excitation of the quantum elements in 3D due to spontaneous emission from nearby neighbours, we have implemented a multi-directional basis for the quantum state of each emitter.
		
		The dynamics of the driven quantum ensemble is characterized by a ``disorder onset rate" that is a function of number density. This ensemble disorder-onset rate reflects the effect of interactions between atoms and, is relatively high for dense, strongly-interacting systems.  The presence of this disorder is immediately significant as it sets an effective time limit in which quantum optical effects are relevant in ensemble dynamics. It also serves as a limit on the applicability of theoretical techniques such as the short-pulse method and simplified basis sets, the use of which may lead to overestimates of coherent effects in quantum ensembles.  
		
		Lastly, we have provided a theoretical method in which the disorder produced during the evolution of a driven  ensemble of quantum emitters can be modelled as decoherence a single particle, specifically, a dephasing.  We have used this model to approximate the state evolution of a dense quantum ensemble using an effective single-particle density matrix.  This method works by allowing for FRET-like coupling between multiple quantum emitters in the ensemble.  This method provides a pretty close approximation to the full, mean-field simulation in significantly less computational time than the full simulation.    This single-particle model also highlights how decoherence processes affect overall ensemble behaviour, which may prove useful in designing protocols for decoherence control.

\bibliography{disp}
\bibliographystyle{unsrt}

\end{document}